\documentclass[11pt]{article}
\usepackage{moriond,epsfig}

\def\be{\begin{equation}}
\def\ee{\end{equation}}
\def\bea{\begin{eqnarray}}
\def\eea{\end{eqnarray}}

\begin{document}
\vspace*{4cm}
\title{DIFFRACTION AT H1 AND ZEUS}

\author{P. J. LAYCOCK}

\address{Department of High Energy Physics, Oliver Lodge Laboratory,\\
University of Liverpool, Liverpool, L69 7ZE, UK}

\maketitle\abstracts{ The H1 and Zeus collaborations have measured the
  inclusive diffractive DIS cross section $ep \rightarrow eXp$ and
  these measurements are in good agreement within a normalisation
  uncertainty.  Diffractive parton density functions (DPDFs) have been
  extracted from NLO QCD fits to these data and the predictions of
  these DPDFs compare well with measurements of diffractive dijets in
  DIS, proving the validity of the factorisation approximations used
  in their extraction.  The inclusive and dijet data are then used in
  a combined fit to constrain the diffractive singlet and gluon with
  good precision over the full phase space.  The predictions of DPDFs
  are compared to diffractive dijets in photoproduction where the
  issue of survival probability in a hadron-hadron environment can be
  studied.  Finally, exclusive diffractive vector meson production and
  deeply virtual Compton scattering have also been studied; the results
  compare reasonably well with the expectations of QCD and in
  particular with GPD models.}

\section{Inclusive Diffraction at HERA}

It has been shown by Collins~\cite{Collins} that the NC diffractive
DIS process $ep\rightarrow eXp$ at HERA factorises; a useful
additional assumption is often made whereby the proton vertex dynamics
factorise from the vertex of the hard scatter - proton vertex
factorisation.  The kinematic variables used to describe inclusive DIS
are the virtuality of the exchanged boson $Q^2$, the Bjorken scaling
variable $x$ and $y$ the inelasticity.  In addition, the kinematic
variables $x_{I\!P}$ and $\beta$ are useful in describing the
diffractive DIS interaction.  $x_{I\!P}$ is the longitudinal
fractional momentum of the proton carried by the diffractive exchange
and $\beta$ is the longitudinal momentum fraction of the struck parton
with respect to the diffractive exchange; $x=x_{I\!P}\beta$.  The data
are discussed in terms of a reduced diffractive cross-section,
$\sigma_r^{D(3)}(\beta, Q^2, x_{I\!P})$, which is related to the
measured differential cross section by:
\begin{equation}
\frac{d^3\sigma_{ep \rightarrow eXp}}{d\beta dQ^2 dx_{I\!P}} = \frac{4\pi\alpha_{em}^2}{\beta Q^4}(1 - y + \frac{y^2}{2})\sigma_r^{D(3)}(\beta, Q^2, x_{I\!P}).
\end{equation}

\begin{figure}[h]
\begin{center}
\includegraphics[width=0.4\columnwidth]{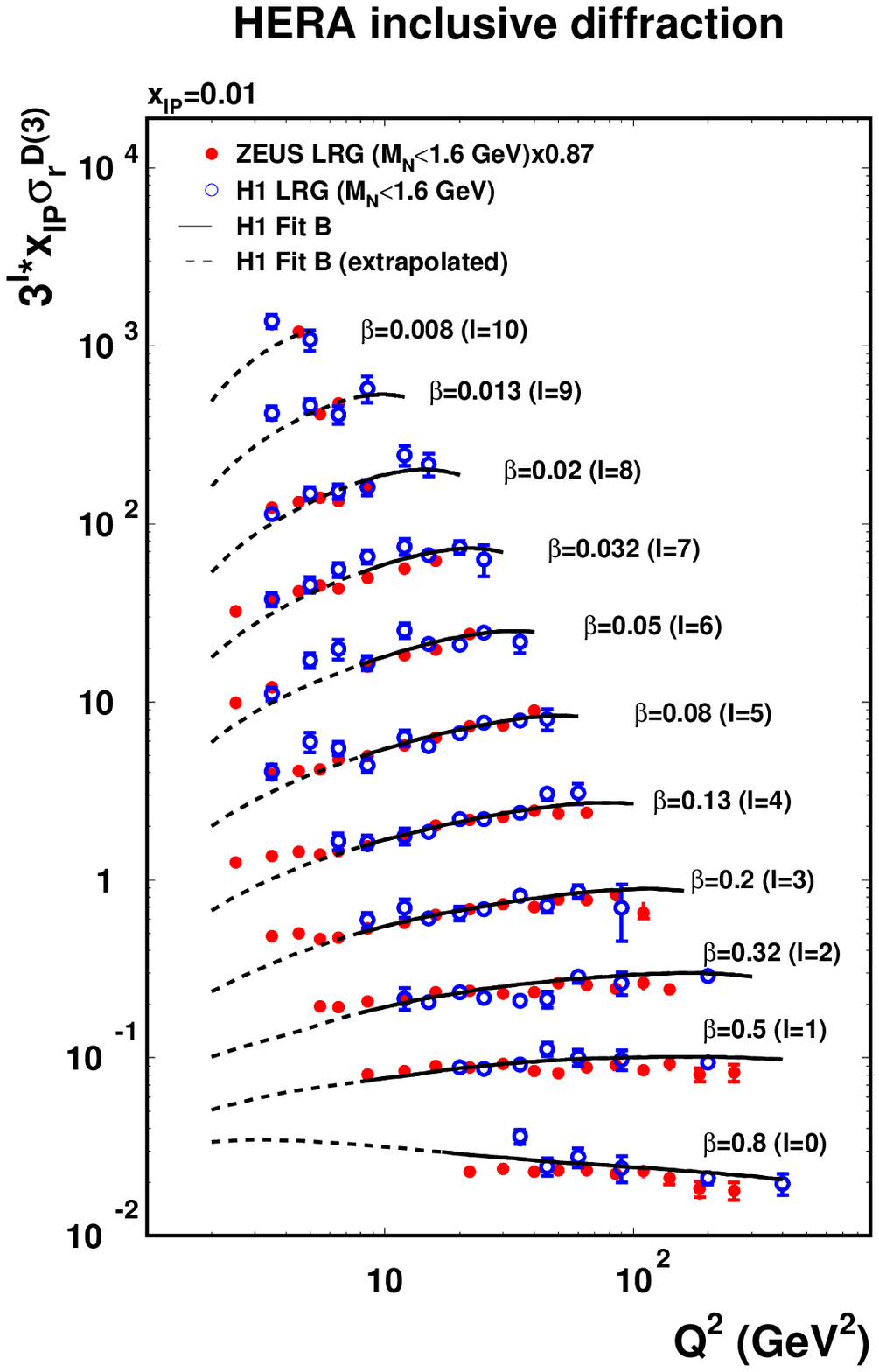}
\includegraphics[width=0.4\columnwidth]{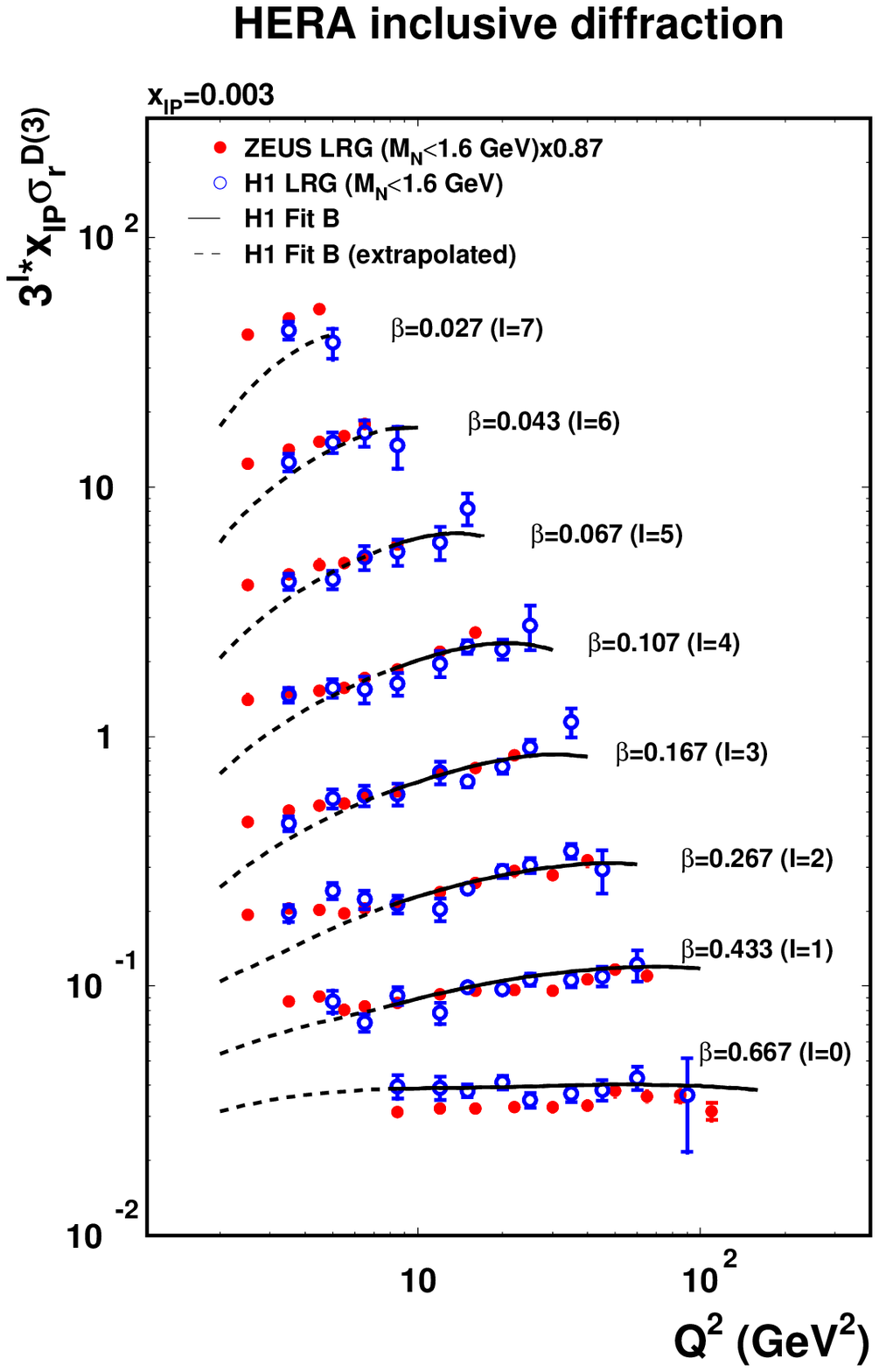}
\caption{The reduced diffractive cross section as measured by the H1 and Zeus collaborations.}
\label{Fig:sigma}
\end{center}
\end{figure}

In the proton vertex factorisation scheme, the $Q^2$ and $\beta$
dependences of the reduced cross section factorise from the $x_{I\!P}$
dependence.  Measurements of the reduced diffractive cross section
from both H1 and Zeus are shown in Figure~$\ref{Fig:sigma}$, where the
new Zeus preliminary measurement has been scaled by a factor of 0.87,
a factor consistent with the normalisation uncertainties of the two
analyses.  The measurements agree rather well.

\subsection{Diffractive PDFs from Inclusive data}

Using the approximation of proton vertex factorisation, the
H1~\cite{H1Inc} and Zeus~\cite{Zeus} collaborations have extracted
DPDFs using NLO QCD fits to the $\beta$ and $Q^2$ dependencies of the
reduced cross section.  H1 obtained two fits of approximately equal
quality, Fit A and Fit B, differing only in the number of terms used
to parameterise the gluon.  The two fits, while fully consistent at
low fractional momentum, yield very different results for the
diffractive gluon at high fractional momentum.  This is due to
quark-driven evolution dominating the logarithmic $Q^2$ derivative of
the reduced cross section at high $\beta$, which in turn greatly
reduces the sensitivity of this quantity to the gluon.
\subsection{Diffractive dijets and DPDFs}

Diffractive dijets in DIS provide a sensitive experimental probe of
the diffractive gluon, as the dominant production mechanism is
boson-gluon fusion.  The sensitive variable is $z_{I\!P}=\frac{Q^2 +
  M_{12}^2}{Q^2+M_X^2}$, where $M_{12}$ is the invariant mass of the
dijet system and $M_X$ is the invariant mass of the total hadronic
final state $X$.  Both H1~\cite{Dijets} and Zeus~\cite{Zeus_Dijets}
have measured the diffractive dijet cross section in DIS.  Both
collaborations find that, at low $z_{I\!P}$, where the inclusive data
have sensitivity to the diffractive gluon, the results of the
predictions are very similar and agree well with the data.  This
supports the use of the proton vertex factorisation approximation
needed to make the NLO QCD fits.  At high $z_{I\!P}$ the data clearly
prefer the prediction of Fit B.

\begin{figure}[h]
\begin{center}
\hspace{-1.0cm}
\includegraphics[width=0.15\textwidth]{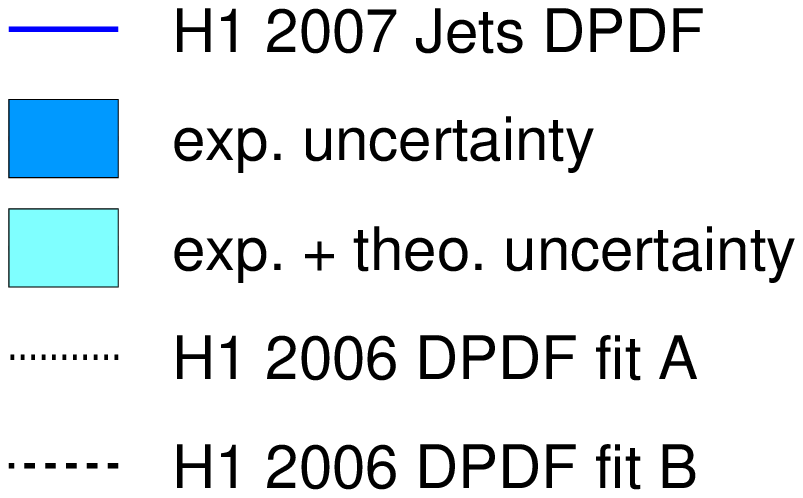}
\hspace{12.0cm}
\vspace{0.1cm}
\includegraphics[width=0.25\textwidth]{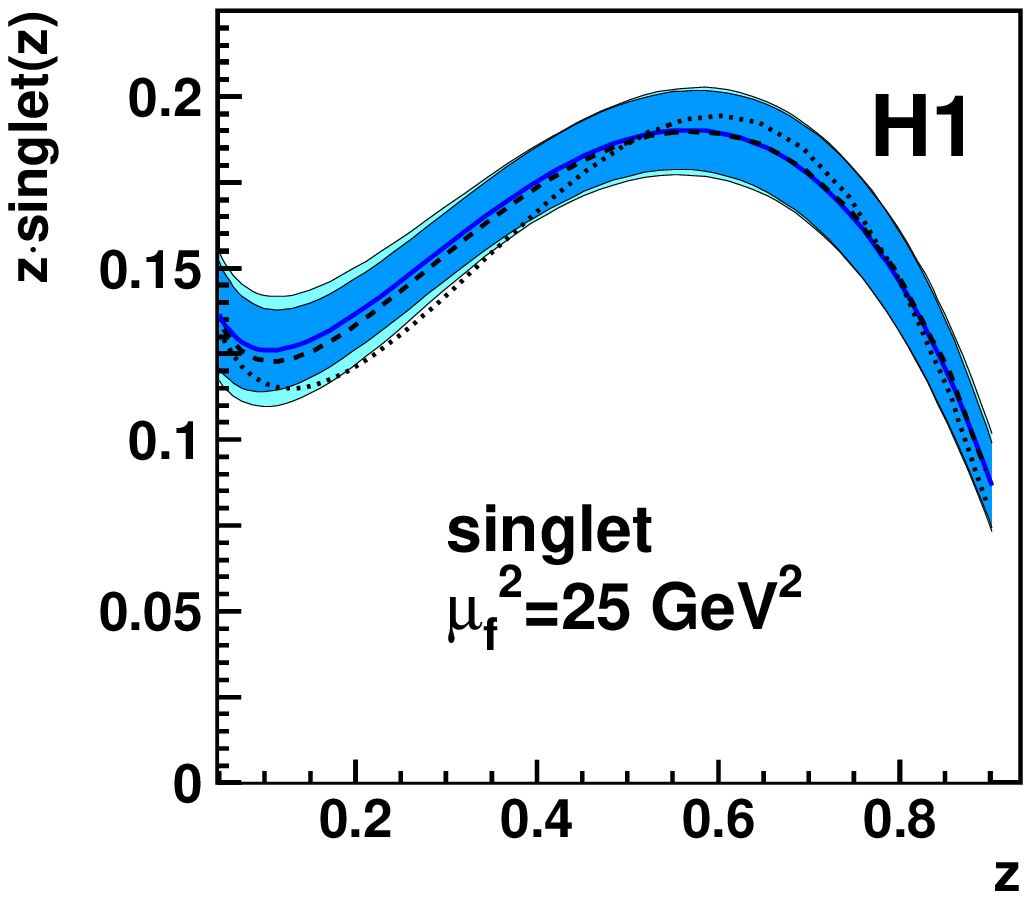}
\includegraphics[width=0.25\textwidth]{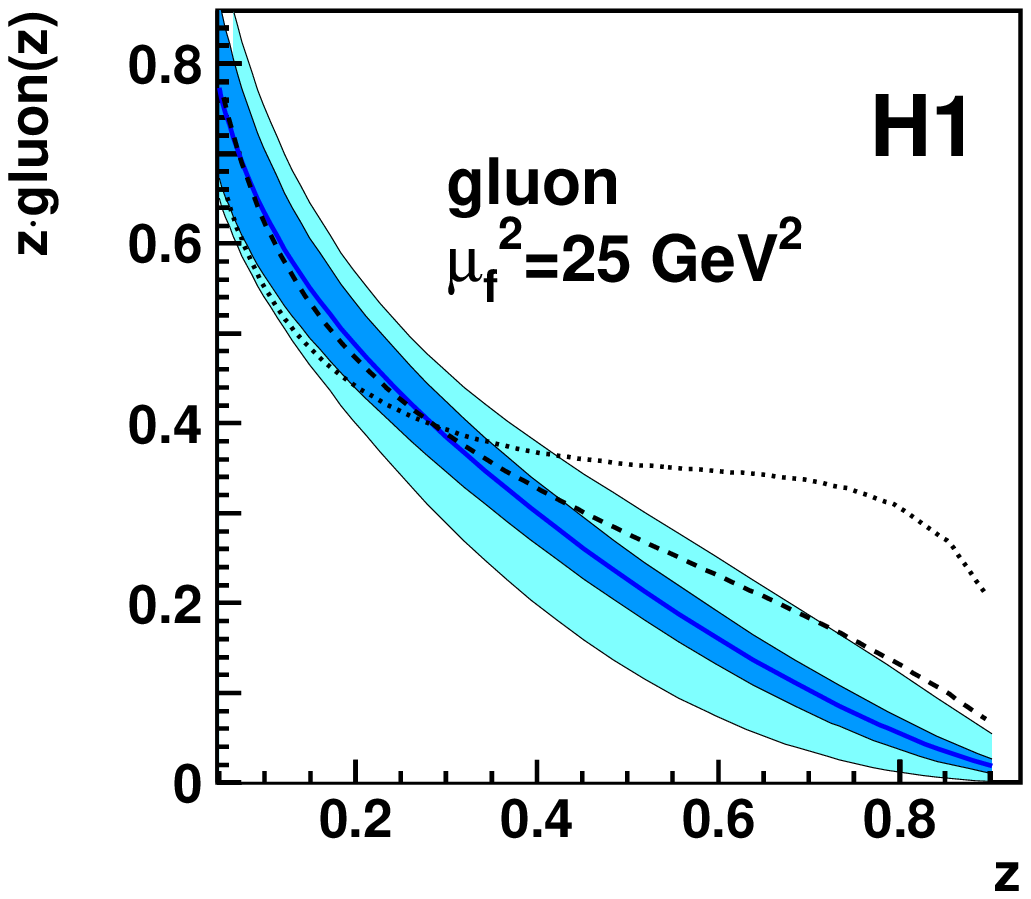}
\hspace{5.0cm}
\vspace{0.1cm}
\includegraphics[width=0.25\textwidth]{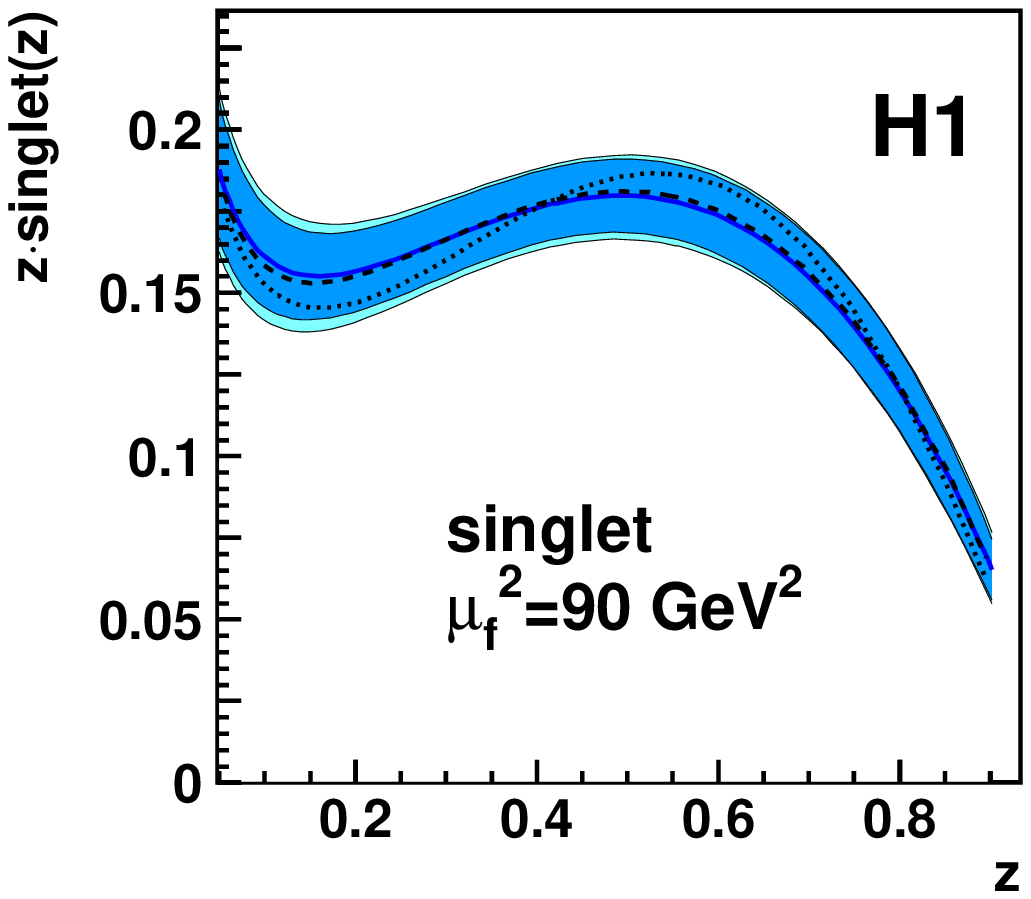}
\includegraphics[width=0.25\textwidth]{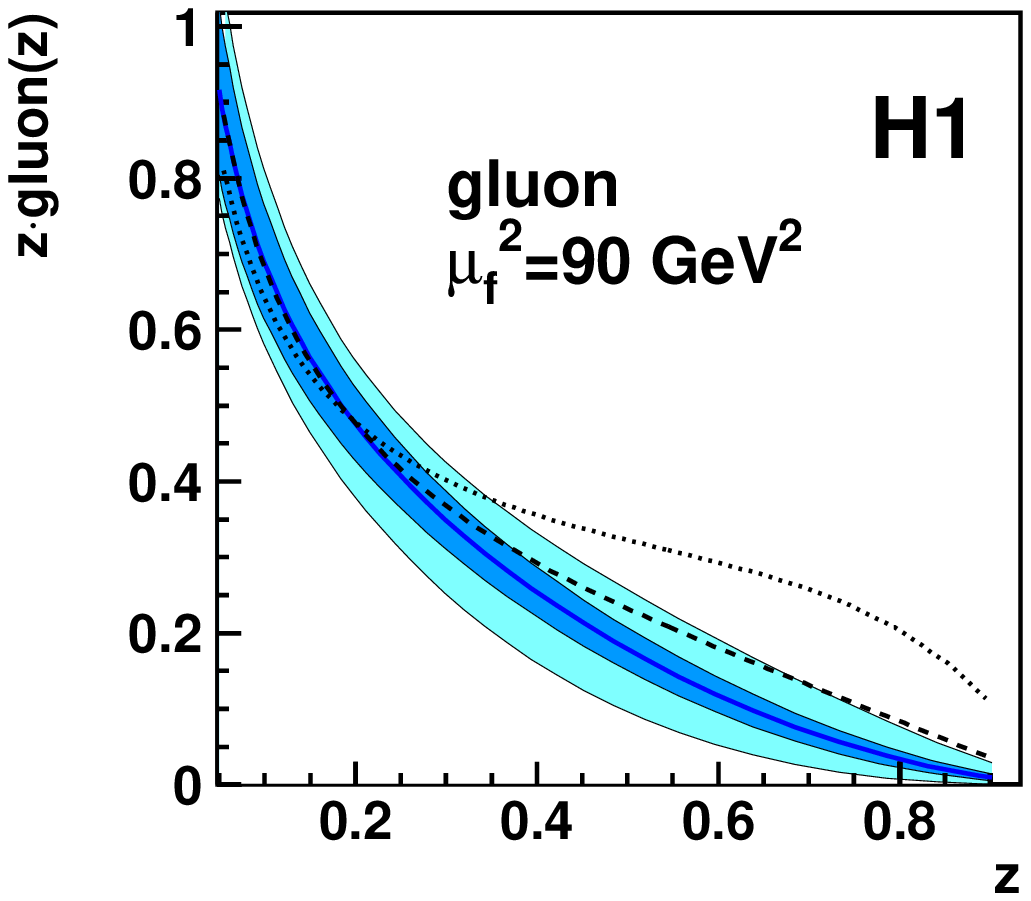}
\end{center}
\caption{The H1 DPDFs resulting from 
the combined fit to the inclusive and dijet diffractive DIS data.}
\label{Fig:JetDPDFs}
\end{figure}

Having shown the sensitivity of the diffractive dijets in DIS data, H1
have included their data in a combined fit with the inclusive
diffractive DIS data~\cite{Dijets}.  The resulting fit is
indistinguishable from Fit A and Fit B in its description of the
inclusive data and produces a better description of the diffractive
dijet data, consistent with that of Fit B.  The resulting DPDFs from
this combined fit, are shown in Figure~$\ref{Fig:JetDPDFs}$.  Both
singlet and gluon are constrained with similarly good precision across
the whole kinematic range.  A parametrisation of the DPDFs is
publically available~\cite{DPDFJets}.

\subsection{Diffractive dijets in photoproduction}

Despite the success of factorisation in diffractive DIS at the HERA
experiments, there is a long-standing issue that the predictions
obtained with HERA DPDFs grossly overshoot the diffractive dijet cross
section at the Tevatron.  At HERA, photoproduction events, where
$Q^2\sim0$, provides an environment similar to a hadron-hadron
collider.  The variable $x_{\gamma}$ is the fraction of the four
momentum of the photon transferred to the hard interaction; the lower
the value of $x_{\gamma}$ the more hadron-like the photon.  Both
H1~\cite{Sebastian} and Zeus~\cite{Zeus_dijetsingammap} have measured
diffractive dijets in photoproduction.  The latest preliminary results
from H1~\cite{Karel} show a suppression of the cross section with
respect to the predictions and this suppression is independent of
$x_{\gamma}$.  There is also a suggestion that this suppression is
dependent on the $E_T$ of the jet.  This would be consistent with the
Zeus analysis at higher $E_T$ where less suppression is observed.  It
should be noted in addition that the current measurements have large
experimental and theoretical uncertainties.

\section{Exclusive vector meson production and DVCS}

\begin{figure}[h]
\begin{center}
\includegraphics[width=0.32\columnwidth]{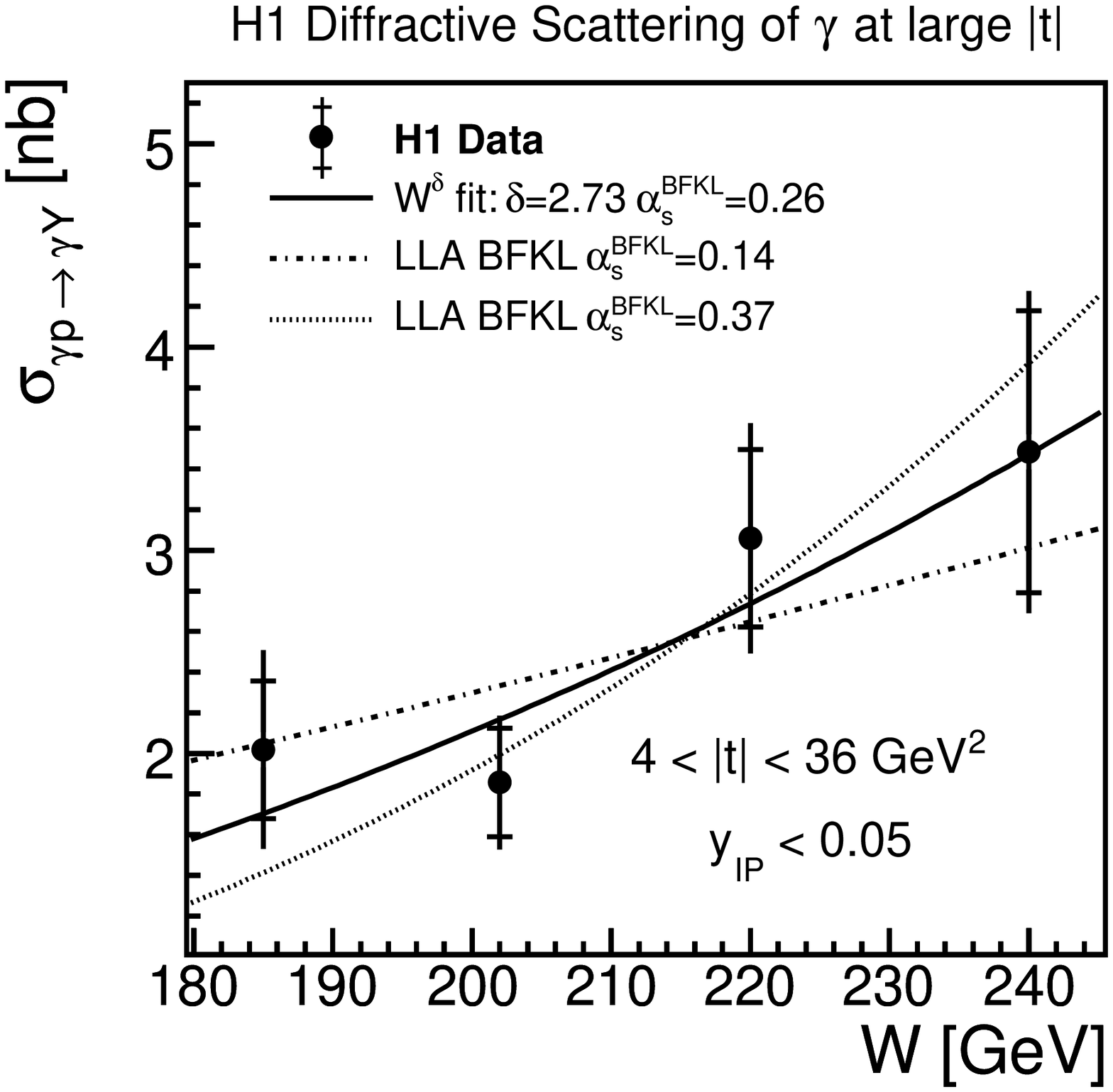}
\includegraphics[width=0.32\columnwidth]{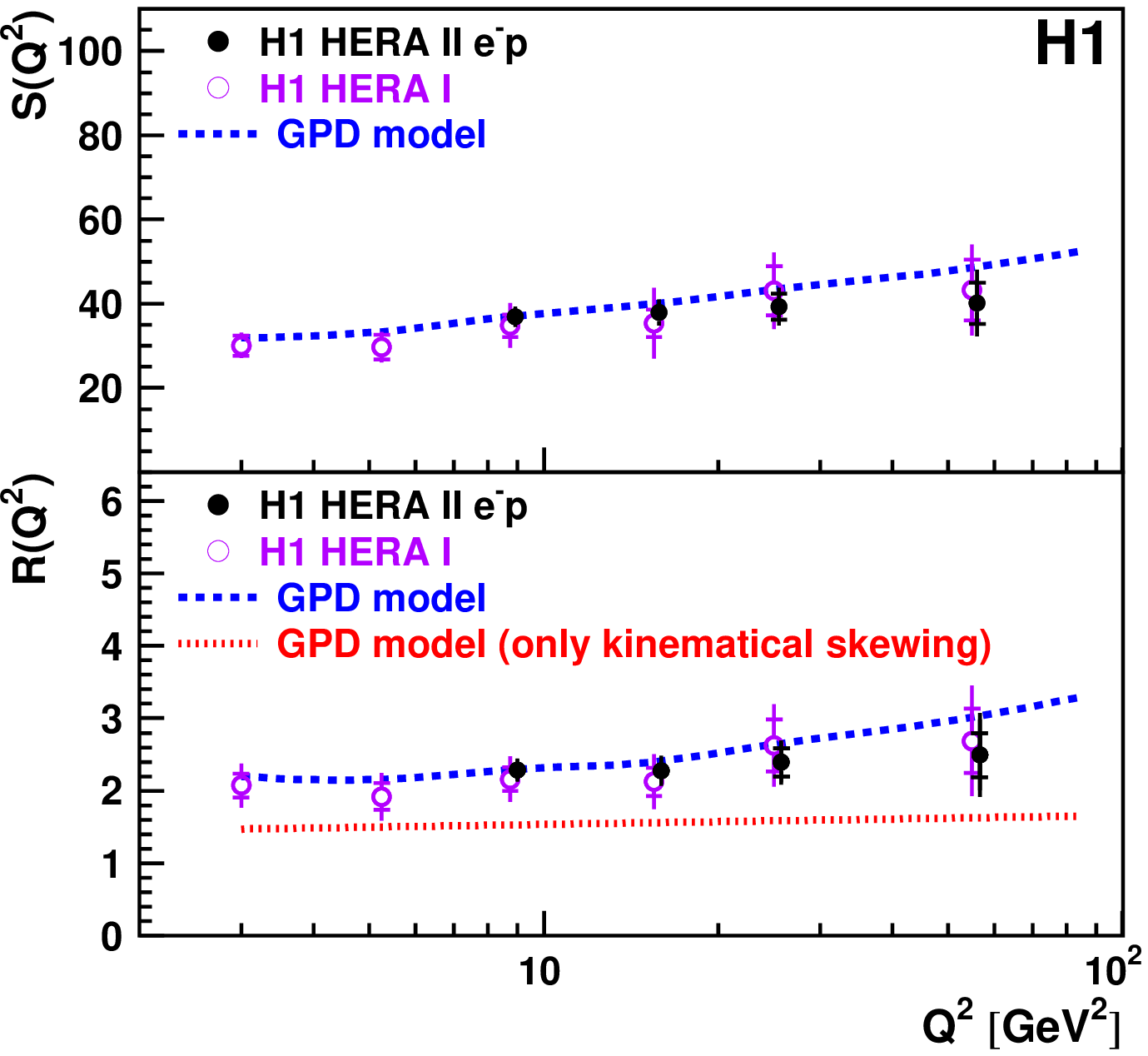}
\caption{The $W$ dependence of the high-$t$ photon cross-section
  (left) and (right) the $Q^2$ dependence of quantities sensitive to
  GPD models (see text).}
\label{Fig:vm}
\end{center}
\end{figure}

Exclusive vector meson production provides an ideal experimental
testing ground for QCD, as the experimental signature is clean and the
theoretical calculations are often simplified.  Measured at
H1~\cite{HighT}, the exclusive production of photons at high momentum
transfer $t$ at the proton vertex allows comparison of the
experimental results with BFKL calculations which do not suffer from
uncertainty on the final state vector meson wave function.  The $W$
dependence of the high-$t$ photon cross-section is shown in figure
$\ref{Fig:vm}$ (left); this is certainly one of the hardest
diffractive processes yet measured and is consistent with the BFKL
predictions, although the precision of the data is limited.

Deeply virtual Compton scattering is a process with sensitivity to the
transverse correlations of partons in the proton and thus has
sensitivity to models of Generalised Parton Densities
(GPDs)~\cite{DVCS}.  Figure $\ref{Fig:vm}$ (right) shows the $Q^2$
dependence of (top) a dimensionless variable $S$ related to the
amplitude for the process with the $t$-dependence removed; (bottom)
the $Q^2$ dependence of a variable $R$ related to the ratio of GPD to
PDF.  The data can discriminate between GPD models and favour a full
GPD model rather than one with only kinematical skewing.

\section{Conclusions}
The H1 and Zeus collaborations have measured the inclusive diffractive
DIS cross section $ep \rightarrow eXp$ and these measurements are in
good agreement within their normalisation uncertainties.  The DPDFs
from NLO QCD fits to the inclusive data can successfully describe
diffractive dijet data in the DIS regime and including these dijet
data in a further NLO QCD fit results in DPDFs constrained with good
precision across the whole kinematic range.  Comparing the predictions
of DPDFs with diffractive dijets in photoproduction shows evidence of
a suppression of the cross section which is independent of
$x_{\gamma}$ but which is consistent with an $E_T$ dependence.
Exclusive vector meson production has also been studied by both the H1
and Zeus collaborations.  H1 have measured exclusive high-$t$ photon
production, a process with one of the hardest $W$-dependences ever
measured.  Measurements of DVCS have been shown to have enough
sensitivity to discriminate between models of GPDs.

\end{document}